\documentclass{llncs}
\usepackage[utf8]{inputenc}
\usepackage{amsmath}   % provides \text{} (used in Sect. "Problems encountered", l.~240)
\usepackage{graphicx}
\usepackage{subfig}
\usepackage[compatibility=false]{caption}
\usepackage{hyperref}
\usepackage{booktabs}
\usepackage{listings}
\usepackage{microtype}
\usepackage{rotating}
\usepackage{tikz}
\usetikzlibrary{arrows.meta, positioning, calc, fit, backgrounds, shapes.geometric}
\begin{document}

\title{Validating ETCS Data with the B Mathematical Language: An Industrial Pipeline and a Blueprint for LLM Integration}

\author{Thierry Lecomte \and Vincent Germain}
\authorrunning{T. Lecomte and V. Germain}
\institute{CLEARSY, Aix en Provence, France,\\
\email{\{thierry.lecomte, vincent.germain\}@clearsy.com}}

\maketitle

\begin{abstract}
Can large language models participate in the production and validation of ERTMS/ETCS data without undermining the certification arguments required by CENELEC EN~50128/50716? ERTMS/ETCS is a distributed safety-critical system (trackside, onboard, radio-block centre) whose behaviour is parameterised by large volumes of data drawn from the UNISIG Subsets; errors in that data propagate through the distributed architecture. This paper reports the current status of an ongoing industrial research effort at CLEARSY, \emph{ValidAItion}, that bridges the ERTMS Operational Simulator to the CLEARSY Data Solver and applies rules expressed in the B mathematical language to that trackside data. During construction, a large language model (Claude) has authored the rule corpus and the parsers through a Model Context Protocol server; every proposal is adjudicated by the downstream toolchain and by systematic human review, and the toolchain has already rejected a syntactically valid but semantically illegal generated scenario. The contribution is architectural and industrial, not algorithmic: the work combines frameworks already in use at CLEARSY (CLEARSY Data Solver, ERTMS Operational Simulator) with a conversational authoring loop, rather than proposing a new formal method. It is a progress report: rule coverage is growing, the human-review campaign is underway, and the quantitative results will be published separately. The paper argues, on the evidence gathered so far, that formal rules in the mathematical language of B must remain the source of truth, while the language model serves as the fenced assistant in a distributed safety-critical railway system: AI proposes, the formal oracle disposes, the human confirms.
\end{abstract}


%-----------------------------------------------------------------------------
\section{Introduction}
%-----------------------------------------------------------------------------

ERTMS and its ETCS layer are being rolled out across an expanding share of the European network. Baseline~3 Release~2 consolidates the specification under the UNISIG Subsets (026 System Requirements, 036 Eurobalise FFFIS, 044 Euroloop FFFIS, 085 Eurobalise test). Production use of ETCS depends not only on \emph{code} being correct, but on \emph{data} (balise telegrams, route data, speed profiles, linking distances, movement-authority ranges) being correct with respect to thousands of rules distilled from those Subsets. Data errors scale with deployment: every new line re-authors hundreds of kilometres of trackside data under time and budget pressure. Data validation is therefore one of the most scalable industrial applications of formal methods in railway, pursued at CLEARSY since the DTVT\footnote{DTVT denotes a data-validation tool used by Alstom Transport SA; the term is retained here as a historical reference to the CLEARSY work of the 2010s on that family of projects.} work of the 2010s~\cite{lecomte2012dtvt,lecomte2024discorail}.

Three pressures are reshaping this activity in 2026. \emph{First}, the growth of Subset coverage expected from modern projects; Baseline~4 will add more \emph{shall} statements. \emph{Second}, the increasing demand for safety in mainline deployment. Formal data validation has a long industrial track record in Communication-Based Train Control for metro lines (CBTC); its use for mainline ERTMS/ETCS is recent. At the same time, the CENELEC standards applicable to this activity have tightened: the successor of EN~50128, EN~50716, now lists formal methods as \emph{highly recommended} at every Safety Integrity Level, SIL-1 and SIL-2 included, whereas EN~50128 only recommended them for those levels~\cite{cenelec50128,cenelec50716}. Safety authorities have reinforced the same expectation: following the 2019 La Milesse overspeed event on the Paris--Rennes high-speed line~(TGV operating under ETCS~Level~2, travelling at 165\,km/h through a turnout authorised to 100\,km/h), the French BEA-TT recommended that the evaluation of ETCS-data validation make explicit use of verification, testing and formal-proof tools~\cite{beatt2020milesse}. \emph{Third}, the arrival of large language models in engineering workflows: LLMs accelerate drafting and exploration but also produce plausible-looking artefacts that are semantically wrong. In a context further regulated, for AI-assisted systems, by the EU AI Act Article~14~\cite{aiact2024}, an LLM that confidently writes invalid ETCS data is worse than no LLM at all.

\paragraph{Position.} Rather than treat LLMs as a replacement for formal methods, we argue for an inversion: formal rules in the B mathematical language~\cite{Abrial-B-Book} (set theory, first-order logic, relations, arithmetic; no B machines, refinements or implementations) remain the source of truth, and LLMs are fenced assistants: every output is adjudicated by a formal oracle and by systematic human review. This stance follows the broader line sketched at FM~2026~\cite{lecomte2026fmai}; the present paper instantiates it on ETCS Subset-026/036/044/085 trackside data validation.

\paragraph{Scope.} We call this effort \emph{ValidAItion} throughout the paper. It is a progress report. At the time of writing, 22 of the 43 Subset-derived tractable rules are formalised as predicates in the mathematical language of B and further rules are being added; the human-review campaign is underway; and the full 82-rule solver sweep on the three ERTMS Operational Simulator (OP~SIMU) production databases is in preparation. The paper establishes the pipeline, the architectural stance, the coverage denominator and the governance structure; the quantitative outcomes of the review campaign and of the production sweep will appear in a follow-up paper.

Section~\ref{sec:background} gives the ERTMS/ETCS background. Section~\ref{sec:rationale} explains the rationale. Section~\ref{sec:pipeline} describes the ValidAItion pipeline. Section~\ref{sec:results} reports the case study, coverage, rule corpus, error taxonomy and human-review artefacts. Section~\ref{sec:blueprint} presents the blueprint for fenced LLM-assisted engineering. Section~\ref{sec:related} surveys related work. Section~\ref{sec:conclusion} concludes.

%-----------------------------------------------------------------------------
\section{Background: ERTMS/ETCS and the Subsets}\label{sec:background}
%-----------------------------------------------------------------------------

ERTMS is a distributed, layered system. Trackside equipment (balises, radio-infill units, radio-block centres) transmits information to onboard units (OBUs) that compute a Movement Authority (MA), monitor the train's adherence to a Static Speed Profile (SSP) and apply service or emergency braking when necessary. The correctness of this distributed behaviour depends on the consistency of the data that parameterises it: the same balise-group identifiers, linking tables, MA segments and SSP entries are read by the trackside generator, encoded into telegrams, transmitted through the radio-balise protocols, and interpreted by the onboard unit. An inconsistency at any point in that chain manifests as a distributed fault. The data-validation activity addressed in this paper targets this parameter layer of the distributed system. The information flow is standardised by the UNISIG Subsets, which define the radio and balise protocols (Subset-036, Subset-044, Subset-058), the test specifications (Subset-076, Subset-085), and the overall System Requirements Specification (Subset-026).

Among the Subsets, \textbf{Subset-026} carries the bulk of the functional specification: it defines the ETCS modes and transitions, the packets exchanged between trackside and onboard, and the operational rules for levels L0, L1, L2 and L3. \textbf{Subset-036} specifies the Eurobalise FFFIS (physical-layer signalling, telegram coding, CRC), and \textbf{Subset-085} specifies how those telegrams must be tested. \textbf{Subset-044} covers the Euroloop FFFIS. Together they constrain every bit encoded in a balise telegram and every kilometre of track data that a project must author and validate before commissioning.

The rules derived from these Subsets fall into two broad categories that matter for the present work. A first category, which we call \emph{geometric/static}, governs the physical placement and identification of trackside equipment: balise groups (BGs) must be a minimum distance apart, must lie on valid edges, must carry unique identifiers within a project; gradient profiles must cover the extent of each MA; SSP speed entries must be positive and non-decreasing on certain segments; linking distances must lie within ranges specified for each Level. These rules can be checked purely from a geometric model of the track (nodes, edges, signals, balise groups, telegrams' linking tables) without simulating the onboard unit. A second category, which we call \emph{onboard/operational}, governs the reaction of the OBU to received information, the bit-level encoding and scrambling of telegrams, the operational procedures for session handover, and the definition of national values. These rules require either a full model of the onboard state machine, a bit-exact decoder of the telegram payload, or operational knowledge that is orthogonal to the trackside dataset.

This partition is decisive for the scope of any industrial data-validation pipeline built from trackside sources alone. In Sect.~\ref{sec:coverage} we give a quantitative classification along these lines for the 147 rules drawn from Subsets~026, 036, 044 and~085, and we use it as the explicit coverage denominator for our claims.

Two further points bear on the rest of the paper. First, ETCS data validation is by nature \emph{scalable}: the rule set is finite and standardised, while the data it must check grows with every deployed kilometre of line. The rule-formalisation cost is paid once and amortised across many deployments, an economic profile favourable to formal methods. Second, the quality of the realistic data used in validation matters: toy examples do not surface the rule-interaction patterns that arise in production topologies. In our case study we rely on the databases of the ERTMS Operational Simulator (OP~SIMU), which carry industrial-grade track layouts including the nine-track CLEARSY demonstration database.

The remainder of the paper uses the word \emph{fencing} for the architectural stance adopted throughout: an LLM proposes engineering artefacts (rules, parsers, scenarios), and every proposal is filtered by a formal oracle (the CLEARSY Data Solver discharging predicates expressed in the mathematical language of B over JSON, or OP~SIMU on scenario reload) and by systematic human review before it is retained. The metaphor is structural rather than rhetorical: the formal toolchain and the human reviewer act as a wall around the LLM, letting through only what they can adjudicate against a Subset clause or a documented engineering convention. The pipeline described in Sect.~\ref{sec:pipeline} instantiates this stance on ETCS trackside data; the blueprint of Sect.~\ref{sec:blueprint} generalises it to engineer-in-the-loop use.

%-----------------------------------------------------------------------------
\section{Rationale: Why B-Math, Why Data Validation, Why Now}\label{sec:rationale}
%-----------------------------------------------------------------------------

The motivation is to reconcile two activities pursued in largely separate teams at CLEARSY. On one side, the formal modelling of software and systems and the data validation of the constant parameters that parameterise them, grounded in the B method for code and in the B mathematical language for project data~\cite{lecomte2012dtvt,lecomte2020clearsy,10.1007/978-3-030-58298-2_8}. On the other, the simulation of ETCS systems for education, training and testing, integrated with testbenches for onboard and trackside equipment and used to study how change requests to the UNISIG Subsets modify the overall behaviour~\cite{lecomte2024discorail}. These two worlds rarely meet: the first operates on structured data with formal tools, the second on behaviour with simulators. The present work connects them and shows, on real OP~SIMU databases, that formal methods can be applied within a pure simulation activity.

The industrial vehicle of the approach is the \emph{CLEARSY Data Solver}\footnote{The CLEARSY Data Solver (\url{https://www.clearsy.com/en/tools/data-solver/}) is the industrial packaging of the Caval rule engine with the ProB constraint solver as back-end, developed within the Atelier~B ecosystem. The product is certified T2 under CENELEC EN~50128~\cite{cenelec50128}, which makes it usable as a verification tool for safety-related software up to SIL~4. Version details of the underlying engines are given in Sect.~\ref{sec:setup}.}, in which rules are written in a controlled B-math syntax with \emph{intermediate constructions} (\texttt{.ic}), \emph{accessors} (\texttt{.acc}) and \emph{compute variables} (\texttt{.cv}). The choice of the B mathematical language is pragmatic: its core (set theory, relations, first-order logic, arithmetic) is expressive enough for the data-validation obligations drawn from the Subsets, and the solver is a mature industrial platform familiar to the CENELEC certification community. We make no claim of new formalism or new solver technique; the contribution of this paper is architectural and industrial, not algorithmic. The choice of B-math, the Atelier~B ecosystem, OP~SIMU and the CLEARSY Data Solver is not a neutral evaluation of alternatives but reflects the frameworks already in industrial use at CLEARSY today; the research question we address is how to combine them with LLM-assisted authoring, not how to replace them. The rules are predicates over the project data, not software to be executed on a train: no B machine, refinement or implementation is produced in this activity.

The target is OP~SIMU itself. OP~SIMU executes ETCS scenarios on a reference virtual machine and stores each scenario as a MySQL dump; the dumps are the authoritative input of the simulator and the primary source of our test data. ValidAItion converts a dump to the solver's input format, discharges the formal rules derived from Subsets~026/036/044/085, and returns an OK/KO verdict together with counter-examples. The commitment is deliberately scoped: the translators and generators of the chain (SQL-to-JSON bridge, parsers, LLM-authored rule drafts) are not themselves certified today; the formal stance is carried by the solver's discharge of the B-math predicates and by the subsequent human review. The reason the activity deserves attention \emph{now} is the LLM question: by 2026 signalling engineering will not remain untouched by LLM-based assistance, and the proper place for an LLM in a safety-critical ecosystem is \emph{fenced}: a \emph{proposer} whose outputs are adjudicated by a formal oracle and reviewed by a human, never a \emph{decider}.

%-----------------------------------------------------------------------------
\section{The ValidAItion Pipeline}\label{sec:pipeline}
%-----------------------------------------------------------------------------

\begin{figure}[t]
\centering
\begin{tikzpicture}[
  font=\footnotesize,
  node distance=5mm and 8mm,
  lane/.style={rounded corners=3pt, line width=0.4pt, draw=black!40, fill=#1, inner sep=5pt},
  block/.style={draw=black!70, rounded corners=2pt, fill=white, align=center, minimum height=7mm, minimum width=22mm, inner sep=2pt},
  tool/.style={block, fill=black!4, font=\footnotesize\bfseries},
  doc/.style={block, fill=blue!5, font=\ttfamily\scriptsize},
  ext/.style={block, fill=orange!8, font=\footnotesize},
  hd/.style={font=\footnotesize\bfseries, text=black!55},
  flow/.style={-{Stealth[length=1.8mm]}, line width=0.5pt, draw=black!70},
  feedback/.style={-{Stealth[length=1.8mm]}, line width=0.5pt, draw=black!55, dashed},
  review/.style={circle, draw=red!70, fill=red!10, inner sep=0pt, minimum size=3.6mm, font=\tiny\bfseries, text=red!80!black},
]
%---- Column 1: AUTHORING ----
\node[ext] (subsets) {ETCS Subsets\\026 / 036 / 044 / 085};
\node[tool, below=of subsets] (llm) {LLM (Claude)\\\normalfont via FastMCP};
\node[doc, below=of llm] (rules) {.rule\ \ .ic\\.acc\ \ .cv};
\draw[flow] (subsets) -- (llm) node[midway, review, right=1mm] {1};
\draw[flow] (llm) -- (rules) node[midway, review, right=1mm] {2};

%---- Column 2: DATA ----
\node[ext, right=of subsets] (sql) {OP SIMU SQL dump\\\scriptsize L1\_FS, Webinar, CLEARSY\_Demo};
\node[tool, below=of sql] (dp) {DataProcessor\\\normalfont sql2json\\\normalfont + decoder Pkt 5/12/21/27\\[1pt]\scriptsize\itshape\mdseries developed with Claude Code};
\node[doc, below=of dp] (json) {.json\\.jsonsd};
\draw[flow] (sql) -- (dp) node[midway, review, right=1mm] {3};
\draw[flow] (dp) -- (json) node[midway, review, right=1mm] {4};

%---- Column 3: ADJUDICATION (stacked) ----
\node[tool, right=of sql] (caval) {CLEARSY\\Data Solver};
\node[doc, below=of caval] (okko) {OK / KO reports\\+ counter-examples};
\draw[flow] (caval) -- (okko) node[midway, review, right=1mm] {6};

%---- Cross-column arrows feeding the Adjudication lane (routed below
%     the Data lane, entering the south edge of the adj lane).
%     Targets are computed from `okko' (which is already defined at this
%     point) + 2mm below, to reach the lane's south border without
%     forward-referencing the fit node.
\draw[flow] (rules.south) -- ++(0,-18mm) -| ($(okko.south)+(-8mm,-2mm)$);
\draw[flow] (json.south)  -- ++(0,-4mm)  -| ($(okko.south)+(8mm,-2mm)$);

%---- LLM-generated scenario feedback loop to OP SIMU (review 5) ----
\draw[feedback] (llm.east) to[bend left=10] node[midway, review, above=0mm] {5} (sql.west);

%---- Lane backgrounds ----
\begin{scope}[on background layer]
  \node[lane=green!6,  fit=(subsets)(llm)(rules)] (authoring) {};
  \node[lane=blue!6,   fit=(sql)(dp)(json)]       (data)      {};
  \node[lane=orange!6, fit=(caval)(okko)]         (adj)       {};
\end{scope}

%---- Column headings on top ----
\node[hd, above=1mm of authoring.north] {AUTHORING};
\node[hd, above=1mm of data.north]      {DATA};
\node[hd, above=1mm of adj.north]       {ADJUDICATION};

\end{tikzpicture}
\caption{ValidAItion pipeline. Three columns: authoring (LLM + MCP), data transport (DataProcessor; sql2json and the telegram decoder were themselves developed with Claude Code under human review), adjudication (CLEARSY Data Solver). The dashed arrow is the LLM-to-OP~SIMU feedback path for scenario editing. Red-circled numbers mark artefacts subject to systematic human review: (1)~Subset-clause-to-rule mapping; (2)~\texttt{.rule} / \texttt{.ic} / \texttt{.acc} / \texttt{.cv} drafts; (3)~OP~SIMU SQL dump provenance; (4)~\texttt{.json} / \texttt{.jsonsd} produced by \texttt{sql2json}; (5)~LLM-generated scenarios before reload into OP~SIMU; (6)~counter-examples before engineering action.}
\label{fig:pipeline}
\end{figure}
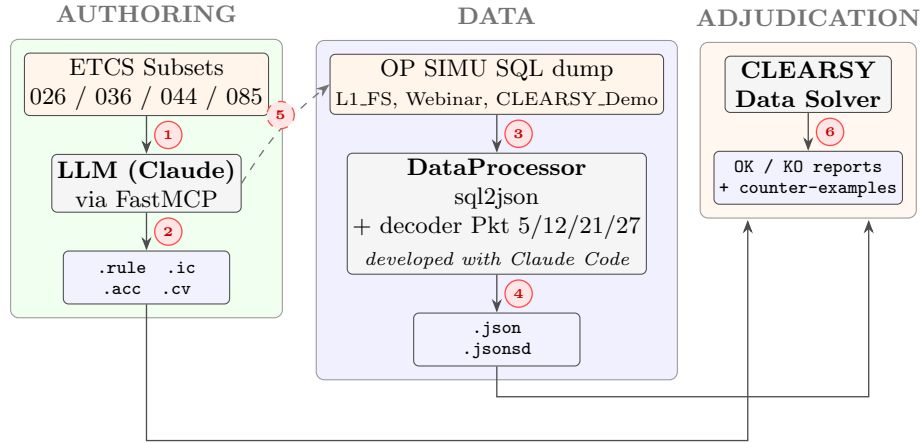

The ValidAItion proof of concept connects two worlds that, until this work, were unconnected inside CLEARSY: the \emph{ERTMS Operational Simulator} (OP~SIMU), which stores ETCS scenarios as MySQL dumps and executes them on a reference VM, and the \emph{CLEARSY Data Solver}, which discharges predicates in the B mathematical language over structured project data. The bridge is a Python command-line interface (CLI), the \emph{DataProcessor}, that parses the OP~SIMU SQL dumps, produces the JSON and JSON-Schema files expected by the solver, carries structured project rules expressed in the mathematical language of B, and orchestrates the execution of those rules. An LLM (Claude), exposed to the database through a Model Context Protocol (MCP) server, participated in the \emph{construction} of the rule corpus and the parsers under human review and was in every case adjudicated by the downstream toolchain. Figure~\ref{fig:pipeline} summarises the three roles of the pipeline (authoring, data transport and adjudication) and marks the six artefacts that are subject to systematic human review.

We emphasise once more that no B machine, refinement or implementation is produced: the rules are predicates over the project data, not software. We make this scope explicit because it has been queried in review. The activity described here intentionally uses the \emph{mathematical language of B} (set theory, first-order logic, relations and arithmetic) as a predicate language over project data; it does not produce B machines, refinements or implementations because the artefact being validated is constant trackside data, not a stateful computation to be refined into code. This is neither a limitation of the B language~---~which can model such a computation when one exists~---~nor a deferral; it is a fit of method to artefact. A future activity that produced a derived safety-critical computation (e.g.\ an onboard runtime check derived from the validated dataset) would justify a chain from B machines through refinement to implementation, and the success rate of such a chain would then be a relevant metric. The present paper does not undertake that step.

\subsection{Architecture, file formats and bridge}\label{sec:arch}

The pipeline has two legs. On the \emph{data leg}, \texttt{dataprocessor sql2json} reads the OP~SIMU MySQL dump, reconstructs the track topology and telegram-linking tables, and emits a \texttt{.json} data file with its matching \texttt{.jsonsd} schema; an integrated decoder handles Packets~5 (linking), 12 (MA), 21 (gradient) and~27 (SSP), while bit-level payload encoding (CRC, scrambling, substitution) remains out of scope. On the \emph{authoring leg}, Claude accesses the same database through a FastMCP server and drafts the rule corpus in six Lark\footnote{Lark, \url{https://github.com/lark-parser/lark}, an LALR/Earley parser library for Python used here to parse the six custom file formats.}-parsed formats: \texttt{.json} data, \texttt{.jsonsd} JSON-Schema, \texttt{.rule} rules in B-math syntax (\texttt{FOR\,/\,WHERE\,/\,SELECT\,/\,VERIFY}), \texttt{.ic} intermediate constructions, \texttt{.acc} accessors over the JSON, and \texttt{.cv} compute variables for auxiliary computations. Round-trip parse/generate is verified by 324 \texttt{pytest} tests; all quantities are in metres with \emph{edge-local abscissae} (positions measured along a track-graph edge, from that edge's start node). Two roles are explicit throughout: the LLM is a \emph{proposer} and the CLEARSY Data Solver is the \emph{adjudicator}; every predicate is discharged by the solver and every LLM-authored edit is confirmed by a human review before being retained.

\subsection{Worked example: \texttt{Rule\_BG\_Min\_Distance}}

The rules are best read in full text. The minimum inter-balise distance within a balise group is governed by SS-036~\S7.5 and specialised at CLEARSY as \emph{ETCS Rule~1.14}; its ValidAItion formalisation is a single \texttt{.rule} file:

\begin{lstlisting}
FOR BAL1, BAL2
WHERE  BAL1 : dom(DEMO::balise_item@NID_BG) & BAL2 : dom(DEMO::balise_item@NID_BG)
     & BAL1 /= BAL2
     & DEMO::balise_item@NID_BG(BAL1) = DEMO::balise_item@NID_BG(BAL2)
     & DEMO::balise_item@edge  (BAL1) = DEMO::balise_item@edge  (BAL2)
THEN SELECT DEMO::balise_item@abscissa(BAL1) <= DEMO::balise_item@abscissa(BAL2)
     THEN VERIFY abs(DEMO::balise_item@abscissa(BAL2)
                   - DEMO::balise_item@abscissa(BAL1)) >= 500
          ENDVERIFY
     ENDSELECT
ENDFOR
\end{lstlisting}

\noindent (The solver scales distances by $10^2$, so 5\,m maps to 500 internal units.) The rule quantifies over two balises of the same balise group on the same \emph{edge} (a track segment between two topological nodes), orders them by abscissa to avoid the symmetric obligation, and asserts their absolute distance is at least 5\,m. The syntax is pure mathematical language of B~\cite{Abrial-B-Book} (set membership, \texttt{dom}, function application, equality, arithmetic) inside a \texttt{FOR\,/\,WHERE\,/\,SELECT\,/\,VERIFY} envelope that the CLEARSY Data Solver compiles into proof obligations. \emph{Accessors} are typed projection functions over the JSON data, defined in the corresponding \texttt{.acc} file; here \texttt{DEMO::balise\_item@NID\_BG} maps a balise identifier to the \texttt{NID\_BG} field of its JSON record. The solver discharges one obligation per pair satisfying the \texttt{WHERE} clause, reports a counter-example on failure, and emits an OK/KO verdict; this rule executes in about seven seconds on a database of the size of the CLEARSY OP~SIMU demonstration scenarios (Sect.~\ref{sec:setup}).

A more complex rule illustrates the predicate language's reach. \emph{Rule\_MA\_Section\_Count\_Valid} (SS-026~\S3.11) asserts that a Movement Authority packet's \texttt{N\_ITER}-encoded section list contains exactly as many sections as declared, that every \texttt{L\_SECTION} is positive and bounded by the project's maximum MA extent, and that the last section's end marker is consistent with the SSP. The rule quantifies over each MA packet in the dataset, projects out the declared and observed section counts via two \texttt{.acc} accessors, traverses the section list with a \texttt{.cv} compute variable, and emits one obligation per inconsistency. Where \emph{Rule\_BG\_Min\_Distance} is a single geometric predicate over a homogeneous collection, this one combines structural traversal, arithmetic bounds and a cross-reference between the MA and the SSP packets; the predicate language handles both without escaping into procedural code.

\subsection{MCP surface and scope invariant}

We call the set of MCP tools exposed to the LLM the \emph{MCP surface}, and we keep this surface narrow by design. The OP~SIMU side exposes a FastMCP\footnote{FastMCP, \url{https://github.com/jlowin/fastmcp}, an open-source Python framework for Model Context Protocol servers.} server with read primitives (\texttt{get\_database\_summary}, \texttt{list\_tables}, \texttt{get\_table\_schema}, \texttt{query\_table}) and create-read-update-delete (CRUD) primitives (\texttt{load\_sql}, \texttt{export\_sql}, \texttt{reset\_database}, \texttt{delete\_entity}, \texttt{update\_entity}), plus further scenario-construction tools. The MCP surface lets the LLM \emph{read} and \emph{propose edits} through a typed, logged API, but never write authoritative artefacts. Those are (i)~the rule corpus expressed in the mathematical language of B, adjudicated by the solver and human-reviewed, and (ii)~the scenario SQL files, adjudicated by OP~SIMU on reload. MCP is the only interface between the LLM and the database~---~in this sense a \emph{narrow waist}, since every interaction passes through it~---~and the toolchain remains the \emph{oracle} that adjudicates every artefact.

The pipeline validates only rules discharged against the JSON track model produced by \texttt{sql2json} and against the decoded Packets~5/12/21/27. This scope is bounded by 43 of the 147 catalogued Subset rules. We have authored 82 rules against this scope: 22 Subset-derived and 60 engineering and integrity rules (see Sect.~\ref{sec:coverage}). The remaining 104 Subset rules require bit-exact payload decoding, OBU simulation, or operational knowledge, and are out of scope. We state this \emph{scope invariant} explicitly: no claim made here extends beyond the 43-rule scope.

%-----------------------------------------------------------------------------
\section{Case Study and Results}\label{sec:results}
%-----------------------------------------------------------------------------

\subsection{Experimental setup}\label{sec:setup}

The \texttt{sql2json} bridge has been exercised end-to-end on three OP~SIMU reference databases (\texttt{L1\_FS\_Scenario}, \texttt{Webinar\_BL3\_2026} and \texttt{CLEARSY\_\allowbreak Demo\_\allowbreak v360}, nine tracks), producing their \texttt{.json}/\texttt{.jsonsd} pair; a deliberately-seeded KO dataset, \texttt{complex\_\allowbreak station\_\allowbreak ko}, serves as negative control. As of the present writing, the rule corpus consists of 82~\texttt{.rule} files with their supporting \texttt{.ic}, \texttt{.acc} and \texttt{.cv} auxiliaries, authored by Claude under human review. Out of these 82 rules, 22 are directly derived from clauses of the UNISIG Subsets. The remaining 60 are engineering guidelines and internal-integrity rules that do not correspond to a specific Subset clause; if violated, they would lead the simulator either to reject the scenario or to exhibit ambiguous behaviour (for example switch-edge consistency, balise-abscissa non-negativity, platform or waking-zone overlap, joint uniqueness, Subset-36 reserved-value ranges). The solver discharge of the 42~rules authored in the initial batch has been executed end-to-end on the seeded control dataset, and a first sweep of three telegram-content rules on the \texttt{L1\_FS\_Scenario} production database has been completed. The remaining production-database sweep is reserved for the immediate near-term step. The CLEARSY Data Solver is configured with the Caval 2.6.0-rc3 rule engine (standalone mode) and the ProB 1.12.2-fix1 constraint solver as back-end, with default fractional digits~$=2$.

\paragraph{Data provenance and confidentiality.} All data used in this study is non-confidential. The UNISIG Subsets are publicly available from the European Union Agency for Railways; the three OP~SIMU databases are CLEARSY's internal demonstration and webinar scenarios (authored in-house for training and demonstration purposes), not customer project data. The LLM (Claude) was operated through a controlled MCP interface exposing only these demonstration assets: no customer data, proprietary configuration, or safety case from a CLEARSY project was shared with the model at any stage.

\subsection{Coverage of the Subsets}\label{sec:coverage}

The 147 rules extracted from Subsets~026, 036, 044 and~085 are partitioned as shown in Table~\ref{tab:coverage}.

\begin{table}[ht]
\centering
\caption{Classification of the 147 rules; 43 are tractable from the JSON track model produced by \texttt{sql2json}.}
\label{tab:coverage}
\begin{tabular}{lrcp{6.0cm}}
\toprule
Category & Count & Tractable? & Why \\
\midrule
Track geometry     & 43  & \textbf{Yes} & Edges, nodes, signals, balise groups, distances \\
Telegram content   & 45  & No & Requires decoding balise telegram hex (packets) \\
Onboard behaviour  & 30  & No & Needs OBU simulation (modes, braking) \\
Operational        & 19  & No & Procedures, national values, session management \\
Encoding           & 10  & No & Bit-level CRC, scrambling, substitution (SS-036) \\
\midrule
\textbf{Total}     & \textbf{147} & & \textbf{43 tractable ($29.3\%$)} \\
\bottomrule
\end{tabular}
\end{table}

This partition is the coverage denominator used throughout the paper. Of the 43 rules tractable from JSON (Table~\ref{tab:coverage}), 22 are today formalised as \texttt{.rule} files and discharged by the solver; a further 21 tractable rules remain to be authored (notably the cross-edge balise-distance checks that require path-traversal intermediate constructions, and the more complex packet parsing for incremental SSP distances and MA sections). The 82-rule corpus mentioned in Sect.~\ref{sec:setup} is therefore composed of 22 Subset-derived rules (a subset of Table~\ref{tab:coverage}'s 43 tractable rules) plus 60 engineering guidelines and internal-integrity rules outside the catalogue, which govern dataset-internal consistency rather than Subset clauses; 82 -- 22 = 60.

\paragraph{Why the other four classes are excluded.} The 104 non-tractable rules are out of scope for distinct reasons: \textbf{Telegram content} (45) requires bit-level payload decoding beyond our field-level decoder for packets~5/12/21/27; \textbf{Onboard behaviour} (30) requires an executable OBU model, outside a static trackside check; \textbf{Operational} (19) lives outside the trackside dataset (procedures, session handover, national values); \textbf{Encoding} (10) is enforced by the trackside encoder hardware and a separate conformance campaign (SS-036~\S8: CRC, scrambling, substitution). The pipeline adjudicates the \emph{what} of ETCS data; the excluded classes adjudicate the \emph{how} of its transport, interpretation and operation.

\subsection{Validation results}\label{sec:runs}

\paragraph{Seeded control dataset.} On \texttt{complex\_station\_ko}, out of 42~rules applied, 28 reported \textbf{OK}, 7 \textbf{KO} (each KO matching one of the violations deliberately seeded in the control dataset) and 7 \textbf{NONE} (the dataset did not contain any item relevant to the rule; a scenario with no loops cannot be tested against \texttt{Rule\_Loop\_Max\_Length}). The seven KO rules are \texttt{BG\_NPIG\_Order}, \texttt{Balise\_Within\_Edge}, \texttt{Linking\_\allowbreak \{Distance\_Range\_L1,\allowbreak\ Target\_Exists\}}, \texttt{MA\_VMAIN\_Positive}, \texttt{SSP\_Speed\_Positive} and \texttt{Signal\_\allowbreak DangerPoint\_\allowbreak Positive}; the seven NONE rules cover loops, joints, SvL hierarchy, uniqueness names and waking zones. The consolidated report lists 8~counter-examples across the seven KO rules; 43 SIL4 verify statements passed, 0 errors, 0 warnings; total wall clock $\approx 5$\,min, per-rule 7--11\,s. Two observations matter. First, the 42-rule sweep is compatible with an interactive authoring loop. Second, NONE is a first-class outcome: when a rule has no applicable data, the solver says so rather than returning a spurious OK. A rule silent for lack of data is not a rule that has adjudicated anything, which matters for the fencing stance of this paper. The pending sweep on the three production-intent OP~SIMU databases is expected to yield a materially lower KO rate, since those databases are authored by CLEARSY and well-formed by construction. Its value therefore lies less in finding errors and more in measuring wall-clock scaling, proof-obligation counts and NONE-distribution at realistic topological complexity.

\paragraph{First production-database sweep.} A first discharge on \texttt{L1\_FS\_Scenario} (single-track Level~1 Full-Supervision, 3 signals, 3 balise groups, 6 balises) exercised three telegram-content rules (\texttt{Rule\_\allowbreak MA\_\allowbreak EOA\_Or\_LOA}, \texttt{Rule\_\allowbreak MA\_\allowbreak Section\_\allowbreak Count\_\allowbreak Valid}, \texttt{Rule\_\allowbreak SSP\_\allowbreak Has\_\allowbreak End\_\allowbreak Marker}). All three reported \textbf{OK}, 0 counter-examples, 8\,s per rule. This is our first solver run on a production-intent OP~SIMU database; the full 82-rule sweep on the three production databases remains the immediate near-term step.

\paragraph{Problems encountered while developing the pipeline.} Three classes of issue surfaced during pipeline development; their resolution shaped the final design described in Sect.~\ref{sec:pipeline}.

\emph{JSONSD-driven typing and scaling:} the solver infers types from the JSONSD (e.g.\ \texttt{NID\_BG} is \texttt{string}, not \texttt{integer}) and multiplies \texttt{number}-typed fields by $10^{\text{default\_fractional\_digits}}$; early drafts of distance rules used the wrong type or unscaled values and failed until aligned with the schema.

\emph{Referential integrity enforced at load time:} the \texttt{referencedTypes} field causes the solver to reject data with dangling references before any rule runs. This made ``balise references a valid edge''-style rules redundant and required that even our deliberately-KO test tracks remain loadable (otherwise a test track breaking referential integrity would be rejected before reaching the rule under test).

\emph{OP~SIMU MD5 fingerprints:} two MD5 fields are computed via the \texttt{genmd5} utility on the Red~Hat Enterprise Linux (RHEL)~7 reference VM; LLM-generated scenarios that omit this step are rejected by OP~SIMU on reload, which is the documented fence event of Sect.~\ref{sec:rex}.

Each issue was caught by the solver, by the OP~SIMU loader, or by human review, and logged in the project's internal record.

\subsection{Error taxonomy}\label{sec:taxonomy}

Five durable classes \emph{of error mode} emerge from the KO verdicts on the \texttt{complex\_station\_ko} control, each anchored to a Subset clause and at least one KO rule: \textbf{E1~Balise-group geometry} (SS-036~\S7.5, SS-026~\S3.4; \texttt{BG\_NPIG\_Order}); \textbf{E2~Positionality} (\S3.6; \texttt{Balise\_Within\_Edge}); \textbf{E3~Linking inconsistency} (\S3.4; \texttt{Linking\_\allowbreak Distance\_\allowbreak Range\_L1}, \texttt{Linking\_\allowbreak Target\_\allowbreak Exists}); \textbf{E4~Positivity/range} (\S3.11, \S3.12; \texttt{MA\_VMAIN\_Positive}, \texttt{SSP\_Speed\_Positive}, \texttt{Signal\_\allowbreak DangerPoint\_\allowbreak Positive}); \textbf{E5~Coverage incoherence} (\S3.11, \S3.12; addressed by a rule but not exercised by the control dataset). A sixth class, \emph{uniqueness}, is addressed by at least one rule in the 82-rule corpus but returned NONE on the control. These classes are intentionally \emph{model-independent}, and the 82-rule implementation (22 Subset-derived plus 60 engineering/integrity) is one instance of the partition.

\subsection{Human-in-the-loop review}\label{sec:human-review}

The fencing stance rests on two lines of defence: the solver adjudicates every B-math predicate, and a human reviewer adjudicates every LLM-authored or machine-generated artefact that feeds, shapes or interprets that adjudication. The solver catches what is formally wrong; the reviewer catches what is silently misaligned with intent or with the Subsets. It is important to note that human cross-verification is not an invention of this work: CLEARSY's current industrial data-validation process already requires that rules, intermediate constructions, accessors and the structured input artefacts (JSON, XML when applicable) be reviewed by an engineer other than their designer before being retained. What the present paper adds is the extension of that established cross-verification discipline to LLM-authored artefacts, and its codification into a governance table aligned with EN~50716 tool-qualification expectations. Table~\ref{tab:human-review} makes explicit which artefacts of the chain are reviewed, by whom, against what reference, and what triggers a reject-and-redo cycle.

The review targets fall into three families: \emph{authoring artefacts} produced by the LLM and read into the solver (rules, intermediate constructions, accessors, compute variables, the mapping from Subset clauses to rule identifiers); \emph{transport artefacts} produced by the DataProcessor and OP~SIMU (SQL dumps, JSON/JSONSD pairs, generated scenarios to reload); and \emph{adjudication artefacts} produced by the solver (OK/KO reports, counter-examples, NONE verdicts). Each artefact has a canonical reference: the UNISIG Subsets for rule content, the DataProcessor design documentation for schema conventions, the OP~SIMU simulator itself for scenario loadability, and the project's internal data-validation conventions for engineering-guideline rules. The eight review targets are not ad hoc: they emerge from the three families above with one target per artefact class that materially carries the fencing argument, and together they cover every artefact in the pipeline of Fig.~\ref{fig:pipeline} whose review affects an oracle verdict or its interpretation. Table~\ref{tab:human-review} lists those eight targets with their reviewer, reference and rejection trigger.

\begin{sidewaystable}
\centering
\caption{Human-review targets in the ValidAItion chain. Results columns are populated in a follow-up study.}
\label{tab:human-review}
\footnotesize
\renewcommand{\arraystretch}{1.2}
\begin{tabular}{|p{0.04\textheight}|p{0.30\textheight}|p{0.18\textheight}|p{0.24\textheight}|p{0.24\textheight}|}
\hline
\textbf{\#} & \textbf{Artefact under review} & \textbf{Reviewer} & \textbf{Reference} & \textbf{Rejection trigger} \\
\hline
H1 & Subset-clause to rule-identifier mapping (traceability table linking each Subset clause in scope to one or more \texttt{.rule} files) & Signalling engineer; B-math reviewer & UNISIG Subsets (026, 036, 044, 085) & Clause not covered, misinterpreted, or rule covers more than its clause states \\
\hline
H2 & \texttt{.rule} drafts authored by the LLM (\texttt{FOR\,/\,WHERE\,/\,VERIFY} predicates in B-math) & B-math reviewer & Corresponding Subset clause and the project's B-math style guide & Predicate over-approximates or under-approximates the clause; solver type error; unit mismatch \\
\hline
H3 & \texttt{.ic} intermediate constructions and \texttt{.acc} accessors authored by the LLM & B-math reviewer & Reference JSONSD schema; project accessor conventions & Accessor arity wrong; \texttt{.ic} confuses edge-local and global abscissa; type signature mismatches the JSONSD \\
\hline
H4 & \texttt{.cv} compute variables (auxiliary imperative) & B-math reviewer & CAFE/Caval \texttt{.cv} semantics documentation & Unintended side-effect; off-by-one in loop; numeric overflow at scaled units \\
\hline
H5 & OP~SIMU SQL dump received from the simulator (provenance) & Signalling engineer & CLEARSY in-house demonstration-scenario catalogue & Unknown origin; customer data inadvertently included; MD5 fingerprints missing \\
\hline
H6 & \texttt{.json} / \texttt{.jsonsd} produced by \texttt{sql2json} (schema and content fidelity) & DataProcessor engineer & SQL dump as authoritative source; DataProcessor design spec & Round-trip mismatch; inferred waking zone disagrees with the SQL; type declared \texttt{integer} where JSONSD expects \texttt{string} \\
\hline
H7 & LLM-generated scenarios proposed for reload into OP~SIMU & Signalling engineer & OP~SIMU on reload (the simulator itself) & OP~SIMU refuses to load; semantic anomaly visible in a simulator run \\
\hline
H8 & Solver counter-examples (KO rule reports) before engineering action & B-math reviewer; signalling engineer & The rule under evaluation; the Subset clause it encodes & Counter-example reflects a rule defect rather than a data defect; localisation ambiguous; engineering remediation not derivable \\
\hline
\end{tabular}
\end{sidewaystable}

The table is the governance spine of an EN~50716-compatible tool-qualification argument for the pipeline: every LLM-authored artefact has a reviewer and a reject-and-redo loop before it is retained, and every oracle verdict is human-adjudicated before an engineering action follows. The human-review campaign is ongoing at the time of writing; the quantitative results (counts of accepted, rejected and reworked artefacts per review target; typical rework time; reviewer effort per authored rule; most frequent rejection triggers observed across the eight targets) are deferred to a dedicated follow-up paper, where they will be reported together with the outcome of the full 82-rule solver sweep on the three OP~SIMU production databases. The present paper establishes the governance structure; the quantitative evidence that populates it comes next.

\subsection{LLM-in-construction: return of experience}\label{sec:rex}

The LLM accelerated ingestion of the Subsets into the 147-rule catalogue, the drafting of \texttt{.rule} and \texttt{.ic} artefacts from Subset clauses, the Lark grammars, and the SQL schema synthesis feeding \texttt{sql2json}. It also produced plausible-but-wrong artefacts caught by the toolchain or by a reviewer (wrong accessor arity, edge-local vs global abscissa confusion, over-approximated BG-distance predicates). The archetypal fence event is a scenario SQL file authored by Claude that was syntactically well-formed, parsed correctly, passed the round-trip, yet OP~SIMU refused to load it: the LLM produced a plausible artefact, the oracle rejected it, and no invalid data entered the project. The pragmatic lesson is that an LLM is useful for \emph{drafting} and \emph{exploration}, not a substitute for the toolchain's adjudication and for systematic human review.

%-----------------------------------------------------------------------------
\section{Blueprint: Fencing AI with B via MCP}\label{sec:blueprint}
%-----------------------------------------------------------------------------

Sections~\ref{sec:pipeline} and~\ref{sec:results} described the LLM+MCP+oracle loop used to \emph{build} ValidAItion. The same architecture is a natural substrate for the \emph{use} phase: a signalling engineer who is not a formal-methods expert asks natural-language questions (``which balise groups violate the minimum-distance rule?'', ``propose a gradient adjustment for MA extent over route~R17''), the LLM consults the database through the MCP server and proposes a concrete edit, and the solver discharges the affected rules before the edit is committed. The diagnostic returns to the engineer in natural language, grounded in a structured solver report.

Beyond what exists today (the nine or more create-read-update-delete (CRUD) tools on OP~SIMU databases mentioned in Sect.~\ref{sec:rex}, the LLM-authored rule corpus, and one documented fence event), the blueprint adds a bidirectional tool surface (\texttt{invoke\_\allowbreak caval\_\allowbreak validation(rule\_id,\,scope)} and \texttt{explain\_ko(report\_id)} so the LLM can drive the oracle and reason over its diagnostics), an orchestration pattern (LLM plans, oracle adjudicates, engineer confirms), and a traceability log pairing every LLM proposal with its oracle verdict, which is the basic artefact for controlled AI assistance under EN~50716~\cite{cenelec50716} tool qualification and AI Act Article~14~\cite{aiact2024}. Residual risks are coverage holes (an unformalised rule cannot fence an LLM error in its class, hence the coverage denominator of Sect.~\ref{sec:coverage}), oracle soundness (a wrong rule lets errors through, rule review remains a human responsibility), LLM capability drift, and prompt injection through engineer input. On the CLEARSY Data Solver side, the pattern calls for structured JSON diagnostics, incremental validation (re-discharging only the rules affected by an edit), and a reusable library of \texttt{.ic} / \texttt{.acc} primitives so that rule authoring stays cheap as coverage grows.

%-----------------------------------------------------------------------------
\section{Related Work}\label{sec:related}
%-----------------------------------------------------------------------------

\subsection{Railway formal methods and ETCS}

Closest to our tool stack, Hansen and Leuschel~et~al.\ showed that a formal B model executed by ProB can run at runtime to control real trains in an ETCS Hybrid Level~3 field demonstration with Thales, with per-event response times of 0.03--0.14\,s~\cite{hansen2018etcsl3runtime,hansen2020etcsl3}; Dghaym~et~al.\ develop a parallel Event-B modelling with iUML-B~\cite{dghaym2018iumlbl3}, and Leuschel and Nayeri extend the approach to Level-3 moving block~\cite{leuschel2023etcsl3mb}. These works establish that ProB, the solver back-end used in the CLEARSY Data Solver, is already accepted in safety-critical railway validation contexts. Broader surveys are given in~\cite{fantechi2025sosym,10.1145/3689374}. In the dynamic-behaviour line, Saddem-Yagoubi~et~al.\ use UPPAAL timed automata for the ETCS L3 Loss-of-Train-Integrity scenario~\cite{saddemyagoubi2024lti}. Our work differs in scope from all of these: we do not prove the ETCS protocol itself and we do not target dynamic behaviour; we validate \emph{deployed trackside data} against a rule set drawn from the Subsets, on an industrial simulator database.

\subsection{The B mathematical language for ETCS and railway data}

Within the B tradition, the DTVT work reported in~\cite{lecomte2012dtvt} and the more recent ERTMS simulation effort in~\cite{lecomte2024discorail} are the most immediate predecessors of ValidAItion; both use the B mathematical language to express data-validation predicates rather than B machines to produce software. The CLEARSY Safety Platform report~\cite{lecomte2020clearsy} and the First Twenty-Five Years of Industrial Use of the B-Method~\cite{10.1007/978-3-030-58298-2_8} describe, separately, the industrial host platform that makes the \emph{B method} (machines, refinements and implementations) viable at SIL~4 for safety-critical software, which is a distinct activity. We extend the data-validation line by connecting the CLEARSY Data Solver to the OP~SIMU database through an automatic bridge, and by publishing an explicit coverage denominator over Subsets~026/036/044/085.

\subsection{LLM + formal methods, AI in rail}

LLM-assisted formal methods are an active area in 2024--2026 but remain largely generic. A broader survey of pragmatic AI uses in formal-methods railway projects at CLEARSY (LLM-assisted specification synthesis, AI-assisted B proof tactics, machine-learning perception fenced by a formally verified safety controller~\cite{safety-controller}, and multimodal-LLM transformation of relay diagrams into propositional logic) is given in~\cite{lecomte2026fmai}; the present paper drills into one of those activities, ETCS data validation. Within the ProB / B-method neighbourhood, Vu, Dunkelau and Leuschel validate reinforcement-learning agents with safety shields~\cite{vu2024rlprob}, Gruteser~et~al.\ combine formal models, safety shields and certified control for AI-based train systems~\cite{gruteser2024certified}, and Rossbach~et~al.\ propose an evaluation methodology for AI in autonomous railway systems~\cite{rossbach2024aieval}. Capozucca~et~al.~\cite{capozucca2025chatgptb} and Dixon~et~al.~\cite{dixon2023nnrobot} explore LLM-assisted B pedagogy and neural-network verification respectively. Our distinctive element is that the LLM/MCP loop was used \emph{during construction} of the rule corpus and the parsers, not as an end-user feature, with the CLEARSY Data Solver over B-math predicates as the adjudicating oracle backed by systematic human review.

\subsection{Differentiation against the closest 2025 works}

Two 2025 railway-community papers are closest in framing to the present work and deserve explicit differentiation.

Yar, Idani, Ledru and Collart-Dutilleul~\cite{yar2025bridging} propose an executable-DSL layer over B specifications so that domain experts can animate and validate signalling models without reading B text. The contribution is a \emph{validation interface} for signalling behaviour models. Our work differs on three axes: (i) our target is ETCS \emph{trackside data} (balise groups, linking, SSP, gradients) drawn from an SQL simulator database, not signalling behaviour models; (ii) we use the B mathematical language directly as a predicate language over JSON data rather than animating B specifications; (iii) our engineer-in-the-loop bridge is a conversational LLM+MCP interface, not an executable DSL.

Reiter, Wetenkamp, Schmid, Kretzschmar and Iffl{\"a}nder~\cite{reiter2025toolchain} target the chain from natural-language signalling requirements to formal specifications and then to \emph{verified code}. Their concern is requirement-to-code traceability and software synthesis. Our concern is the complementary problem: we do not synthesise software; we validate \emph{data}, and we do so for ETCS Subset-026/036/044/085 artefacts authored from an existing industrial simulator. Where their toolchain produces code, ours produces an OK/KO verdict over a deployed dataset; both chains need tool qualification under EN~50716~\cite{cenelec50716}, but they qualify different things.

Relative to these two, the distinctive elements of ValidAItion are: the \emph{data-centric} rather than behaviour-centric or code-centric framing, the \emph{SQL-to-B-math} bridge anchored on OP~SIMU, and the \emph{LLM-in-construction} return of experience with a documented fence event (Sect.~\ref{sec:rex}).

%-----------------------------------------------------------------------------
\section{Conclusion}\label{sec:conclusion}
%-----------------------------------------------------------------------------

We have reported on the current status of \emph{ValidAItion}, an ongoing industrial research effort at CLEARSY that connects OP~SIMU to the CLEARSY Data Solver and applies rules expressed in the B mathematical language to ETCS trackside data. The paper makes three contributions: (i)~an industrial pipeline with an automatic SQL-to-JSON bridge, a telegram decoder for Packets~5/12/21/27, and an executable corpus of 82 rules run under the solver (Sect.~\ref{sec:pipeline}, Sect.~\ref{sec:results}); (ii)~a falsifiable coverage classification of the 147 rules drawn from the Subsets, with 43 tractable from the JSON track model (22 authored so far) and 104 out of scope for explicitly justified reasons (Sect.~\ref{sec:coverage}); (iii)~an LLM+MCP construction methodology fenced by the solver and by systematic human review (Sect.~\ref{sec:human-review}), extended to a blueprint for engineer-in-the-loop data validation (Sect.~\ref{sec:blueprint}). A model-independent error taxonomy of five durable classes underpins the empirical spine. One documented fence event illustrates the stance we advocate: OP~SIMU refused a syntactically valid LLM-generated scenario, and no invalid data entered the project. In safety-critical ETCS data work, formal rules in the B mathematical language must remain the source of truth: AI proposes, the formal oracle disposes, the human confirms.

\paragraph{What is done, in progress, and planned.} \emph{Done.} The pipeline is operational end-to-end on the seeded control dataset (42 rules, 28 OK, 7 KO, 7 NONE) and on a first production database (three telegram-content rules on \texttt{L1\_FS\_Scenario}). Rule authoring has reached 82 rules; the coverage denominator, the error taxonomy and the human-review table are established. \emph{In progress.} The human-review campaign is underway on the eight review targets of Table~\ref{tab:human-review}; the full 82-rule solver sweep on the three OP~SIMU production databases is being set up; additional Subset rules are being formalised, in particular the cross-edge balise-distance checks and the incremental SSP / MA-section packet parsing that will close the gap towards the 43 tractable clauses. \emph{Planned.} A follow-up paper will report the quantitative outcome of the human-review campaign (acceptance / rejection / rework counts per target) and of the full solver sweep (per-database OK/KO/NONE distributions, wall-clock scaling, proof-obligation counts), together with the tool-qualification argument for the currently uncertified transport layer under EN~50716. Longer-term work engages with the certification community on the implications of an LLM-assisted but oracle-fenced and human-reviewed data-validation workflow, and characterises the additional human-review effort induced by running a smaller, privacy-preserving LLM on-premises rather than a hosted frontier model. A further research direction concerns \emph{scenario intent}. The current pipeline verifies that a scenario's trackside data complies with the Subsets, but it does not assess whether the scenario achieves the pedagogical or testing objective for which it was designed. For example, a scenario intended to train drivers to handle a telecommunication loss and to use the Override button to resume their journey requires an ETCS Level~2 topology; the same scenario is unrealisable on a Level~1 track, where no radio session exists and Override has no operational meaning. An LLM with sufficient knowledge of ETCS operational modes and of the training curriculum could assist the scenario designer by checking the feasibility of the stated intent against the topology before the scenario is authored in detail, and by flagging mismatches early. This would extend ValidAItion from \emph{data compliance} (are the rules satisfied?) to \emph{intent adequacy} (can this topology support the intended training objective?).

\paragraph{Confidentiality and the private-LLM trade-off.} The present study relies exclusively on non-confidential inputs (public UNISIG Subsets, CLEARSY's in-house demonstration databases), with a hosted frontier LLM behind a narrow MCP surface that never exposed customer data. Real customer projects require running the LLM on private or on-premises infrastructure, as reported in~\cite{lecomte2026fmai}. Open-weight models deployable on private infrastructure (Llama, Mistral, Qwen families) trail hosted frontier models (Claude, GPT, Gemini) by a measurable margin on head-to-head preference evaluations~\cite{chiang2024chatbotarena} and on reasoning, code and instruction-following benchmarks~\cite{liang2023helm}. A fenced architecture partially compensates: the oracle catches a plausible-but-wrong artefact regardless of proposer capability, and the quantitative question becomes not ``can a local LLM be trusted?'' but ``how much human-review effort does a weaker proposer add?'', a natural follow-up study.

%-----------------------------------------------------------------------------
\section*{Acknowledgements}
%-----------------------------------------------------------------------------
The authors thank the CLEARSY data validation team for the industrial context and the OP~SIMU access, and acknowledge the companion work presented at FM~2026~\cite{lecomte2026fmai}.

%-----------------------------------------------------------------------------
% Bibliography
%-----------------------------------------------------------------------------
\bibliographystyle{splncs03}
\bibliography{biblio}

\end{document}